\RequirePackage[svgnames]{xcolor}

\documentclass[twocolumn]{mystyle}

\usepackage[all]{hypcap}
\usepackage[svgnames]{xcolor}

\usepackage[numbers]{natbib}
\usepackage{hyperref}

\usepackage{algorithm}
\usepackage{algorithmicx}
\usepackage{algpseudocode}
\usepackage{microtype}
\usepackage{graphicx}
\expandafter\def\csname ver@subfig.sty\endcsname{}
\usepackage{booktabs} %
\usepackage{float}
\usepackage{bigstrut}

\usepackage{amsmath}
\usepackage{amssymb}
\usepackage{mathtools}
\usepackage{amsthm}
\usepackage{mathrsfs}
\usepackage{nicefrac}
\usepackage{dsfont}
\usepackage{enumitem}
\usepackage{subcaption}
\usepackage{graphicx,subfig}
\usepackage{cleveref}
\usepackage{bxcoloremoji}

\usepackage{float}

\usepackage[utf8]{inputenc} %
\usepackage[T1]{fontenc}    %
\usepackage{hyperref}       %
\usepackage{url}            %
\usepackage{booktabs}       %
\usepackage{amsfonts}       %
\usepackage{nicefrac}       %
\usepackage{microtype}      %
\usepackage{graphicx}
\usepackage{subcaption} 
\usepackage{amssymb}
\usepackage{fdsymbol}
\usepackage{wrapfig}
\usepackage{lipsum}
\usepackage{enumitem}
\usepackage{stackengine}
\usepackage[font=small,labelfont=bf]{caption}
\usepackage{color}
\usepackage{adjustbox}

\usepackage{rotating}
\usepackage{makecell}
\usepackage{xspace}
\usepackage{multirow}

\definecolor{ForestGreen}{RGB}{34,139,34}

\renewcommand{\paragraph}[1]{\medskip\noindent\textbf{#1.~}}

\newcommand{\github}{\raisebox{-1.5pt}{\includegraphics[height=1.05em]{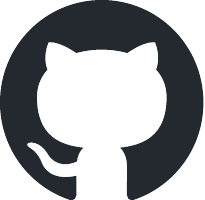}}}

\renewcommand{\paragraph}[1]{\medskip\noindent\textbf{#1.~}}

\newcommand{\modelname}{\texttt{TTTIR}}
\usepackage{amsmath}

\title{\modelname{}: Unlocking Instance-Specific State Evolution via Test-Time Training \\ for Image Restoration}
\runningtitle{ \modelname{}: Unlocking Instance-Specific State Evolution via Test-Time Training for Image Restoration}

\author[1*]{Kaihang Zheng}
\author[2,5*]{\mbox{Jun Li}}
\author[3]{\mbox{Hang Guo}}
\author[5]{\mbox{Hongyu Chi}}
\author[5]{\mbox{Zimo Liu}}
\author[4]{\protect\\\mbox{Tao Dai}}
\author[1\dagger]{\mbox{Jinpeng Wang}}
\author[1,5\dagger]{\mbox{Yaowei Wang}}
\affil[1]{\mbox{Harbin Institute of Technology, Shenzhen}}
\affil[2]{\mbox{Tsinghua University}}
\affil[3]{\mbox{École Polytechnique Fédérale de Lausanne}}
\affil[4]{\mbox{Shenzhen University}}
\affil[5]{\mbox{Peng Cheng Laboratory}}

\authornote[*]{\mbox{Equal Contributions}}
\authornote[\dagger]{\mbox{Corresponding Authors}}

\begin{document}
\begin{abstract}
Image restoration is inherently challenging due to the diverse and highly input-dependent nature of real-world degradations. While recent architectures like Transformers and state-space models have advanced the field, they predominantly rely on static, globally shared parameters, which struggle to fully accommodate instance-specific degradation patterns. Test-Time Training (TTT) offers a promising paradigm for generating data-dependent operators, yet its standard self-supervised inner loop lacks the explicit guidance required to transition degraded features toward clean structures. To address this, we propose \modelname{}, a novel framework that reformulates image restoration as an instance-specific state evolution process. Specifically, we design Progressive State Sequence Generation (PSSG) to construct complementary spatial-frequency target states (defining what to recover), and State Transition Evolution (STE) to adapt lightweight transition operators via a restoration-oriented TTT inner loop (determining how the features should evolve). Extensive experiments demonstrate that \modelname{} consistently outperforms state-of-the-art models across multiple image restoration benchmarks, achieving dynamic instance-specific recovery with favorable computational scalability. 

\vspace{2mm}
\vspace{5mm}

\coloremojicode{1F4C5} \textbf{Date}: September 14, 2026

\github{} \textbf{Code Repository}: \href{https://github.com/Elysiaaaaaaaa/TTTIR.git}{https://github.com/Elysiaaaaaaaa/TTTIR.git}

\coloremojicode{1F4E7} \textbf{Contact}: 
\href{mailto:1023930132zkh@gmail.com}{1023930132zkh@gmail.com} (Kaihang Zheng), \href{mailto:wangjp26@gmail.com}{wangjp26@gmail.com} (Jinpeng Wang)

\end{abstract}

\maketitle
\section{Introduction}
\label{sec: introduction}

Image restoration aims to recover images from observations degraded by diverse factors (e.g., rain, low illumination, haze). Real-world restoration is challenging not only due to task diversity, but also due to inherent \emph{input-dependency}. Even within the same task setting, images may differ in degradation patterns, frequency characteristics, and scene semantics. Therefore, an effective model should determine both \emph{what to recover} and \emph{how to dynamically adapt the restoration process to each input}.

Existing methods like CNNs \cite{chen2022simple}, Transformers \cite{liang2021swinir}, and state-space models \cite{guo2024mambair} have advanced this field through enhanced spatial and sequential modeling. Despite these gains, most models share learned parameters across image instances. Though mechanisms like attention or dynamic gating enable input-specific responses, the underlying mapping functions remain restricted by these static weights. Such shared parameterization may not fully accommodate diverse, input-dependent degradations, motivating a better instance-specific restoration model.

Recently, Test-Time Training (TTT) \cite{sun2024learning} offers a new perspective for input-dependent modeling. Instead of relying on static inference, TTT formulates parameter updates as a dual-level optimization process: an inner loop adapts to the current input and an outer loop learns task-level initialization. By generating instance-specific operators through this input-conditioned parameter update, TTT is suited for modeling instance-level restoration differences.

\begin{figure}[t!]
    \centering
    \includegraphics[width=0.95\linewidth]{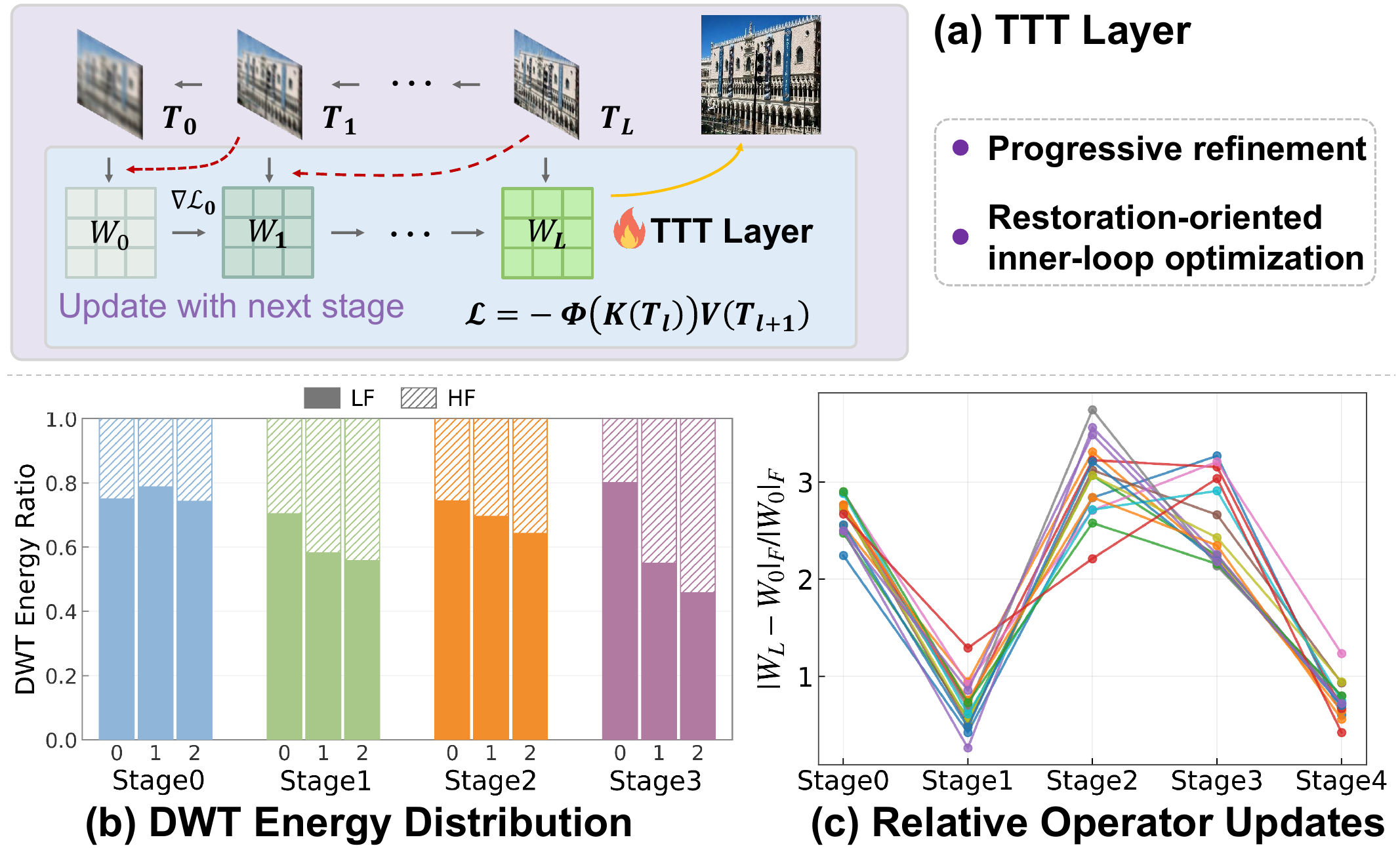}
    \caption{
        (a) The restoration-oriented TTT inner loop of \modelname{}, where the next target state guides input-specific operator adaptation and progressive state transition.
        (b) DWT energy distributions of the target states across different evolution levels.
        (c) Relative magnitudes of inner-loop weight updates induced by different images across restoration stages.
    }
    \label{fig:intro}
\end{figure}
However, the standard TTT layer, primarily designed for 1D sequential data, is sub-optimal for image restoration. In standard TTT, the inner loop acts as a hidden state that compresses historical context via self-supervised updates. While effective for sequence modeling, its inner-loop objective lacks explicit guidance to transition degraded features toward clean representations. Image restoration fundamentally requires degraded features to be progressively guided toward representations with more complete structures and details. This motivates a restoration-oriented inner-loop objective that provides an explicit optimization direction for state transitions.

To address this issue, we propose \modelname{}, which reformulates image restoration as an instance-specific state evolution process. 
\modelname{} consists of Progressive State Sequence Generation (PSSG) and State Transition Evolution (STE). 
PSSG employs a Progressive Spatial State Constructor (PSSC) and a Progressive Frequency State Constructor (PFSC) to generate complementary spatial and frequency states, which are integrated into restoration-oriented target states to guide structural recovery, degradation suppression, and detail reconstruction. 
Guided by these targets, STE employs the restoration-oriented TTT inner loop illustrated in \mbox{\Cref{fig:intro}(a)} to adapt a lightweight transition operator to each input, thereby progressively evolving the current state toward the subsequent restoration state.
Thus, PSSG determines \emph{where} the state should evolve, while STE determines \emph{how} the transition should be performed, enabling input-specific restoration trajectories.

Extensive experiments on 12 benchmark datasets across three restoration tasks demonstrate that \modelname{} achieves state-of-the-art or competitive performance with favorable computational scalability. As shown in \mbox{\Cref{fig:intro}(b)}, frequency-domain analysis reveals a progressive transition from low-frequency structural recovery to high-frequency detail reconstruction. Meanwhile, the operator adaptation analysis in \mbox{\Cref{fig:intro}(c)} shows that different inputs induce distinct parameter-update trajectories. Together, these observations validate the progressive and input-dependent restoration dynamics of \modelname{} from both feature and operator perspectives.

To summarize, we make the following contributions:
\setlist{nolistsep}
\begin{itemize}[leftmargin=1em]
\item We reinterpret TTT as an optimization-based mechanism for generating data-dependent restoration operators, enabling instance-specific computation beyond globally shared restoration mappings.

\item We propose a restoration-oriented state evolution framework that constructs spatial-frequency target states and adapts lightweight transition operators through inner-loop optimization, thereby guiding structural recovery and detail reconstruction.

\item Extensive experiments on multiple image restoration benchmarks validate \modelname{}’s superiority to state-of-the-art. Detailed analyses further justify our key designs.
\end{itemize}
\section{Related Works}
\label{sec:related_work}

\subsection{Image Restoration}

Recent image restoration methods based on Transformer architectures
\cite{he2025universal} and state-space models (SSMs)
\cite{lin2025eamamba, zhang2026beyond} have achieved progress in image restoration tasks.
Despite their architectural differences, these approaches rely on sequential representations constructed from two-dimensional visual features for dependency modeling.

Transformer-based image restoration methods leverage self-attention to capture global dependencies among visual tokens.
However, the quadratic computational and memory complexity of self-attention
\cite{dosovitskiy2020image,liu2021swin}
limits its scalability for high-resolution restoration.
Recent studies alleviate this limitation through efficient attention designs and token interaction strategies
\cite{he2025universal,hu2025enhancing,zhang2026atd},
but often rely on predefined attention patterns or token compression, which may limit flexible global feature modeling.

SSM-based methods achieve efficient sequence modeling with linear complexity through recursive state updates, offering advantages in computational efficiency.
However, due to the causal propagation mechanism and long-range state decay characteristics
\cite{lin2025eamamba,hassanin2026progressive,yoshimura2026sf}, the sequential modeling paradigm of SSMs faces challenges in capturing spatially consistent global dependencies.
Although extensions have been proposed to alleviate these limitations, these methods inherit the structural mismatch introduced by spatial sequence serialization.

\subsection{Test-Time Training}

Test-time adaptation mitigates distribution shifts by optimizing pretrained models on unlabeled test samples via self-supervised objectives. While widely applied to tasks like masked autoencoding \cite{gandelsman2022testtime}, inpainting \cite{ghiro2022test}, dehazing \cite{liu2022towards}, super-resolution \cite{li2024test}, and denoising \cite{mansour2024ttt}, these methods inherently require an explicit adaptation stage for each test instance or target distribution.

In contrast, TTT layers \cite{sun2024learning} formulate sequence modeling as online learning, using models whose fast weights encode context through inner-loop updates within the forward pass. This paradigm has been extended to long-context modeling \cite{zhang2025test}, vision \cite{han2026vit}, video generation \cite{dalal2025one}, spatial intelligence \cite{liu2026spatial}, and 3D reconstruction \cite{jin2026zipmap,wang2026tttlrm}. In low-level vision, MoiréXNet \cite{li2025moirexnet} adopts linear-attention TTT modules for efficient demoiréing.

Unlike TTT layers mainly capturing contextual relationships, \modelname{} constructs spatial--frequency target states and reformulates the inner-loop objective as a restoration-oriented state transition. The targets define the direction, while input-conditioned optimization adapts the transition operator for each image, enabling progressive degradation removal and detail reconstruction.
 \section{Method}
\label{sec:method}

\begin{figure*}[t]
    \centering
    \includegraphics[width=\textwidth]{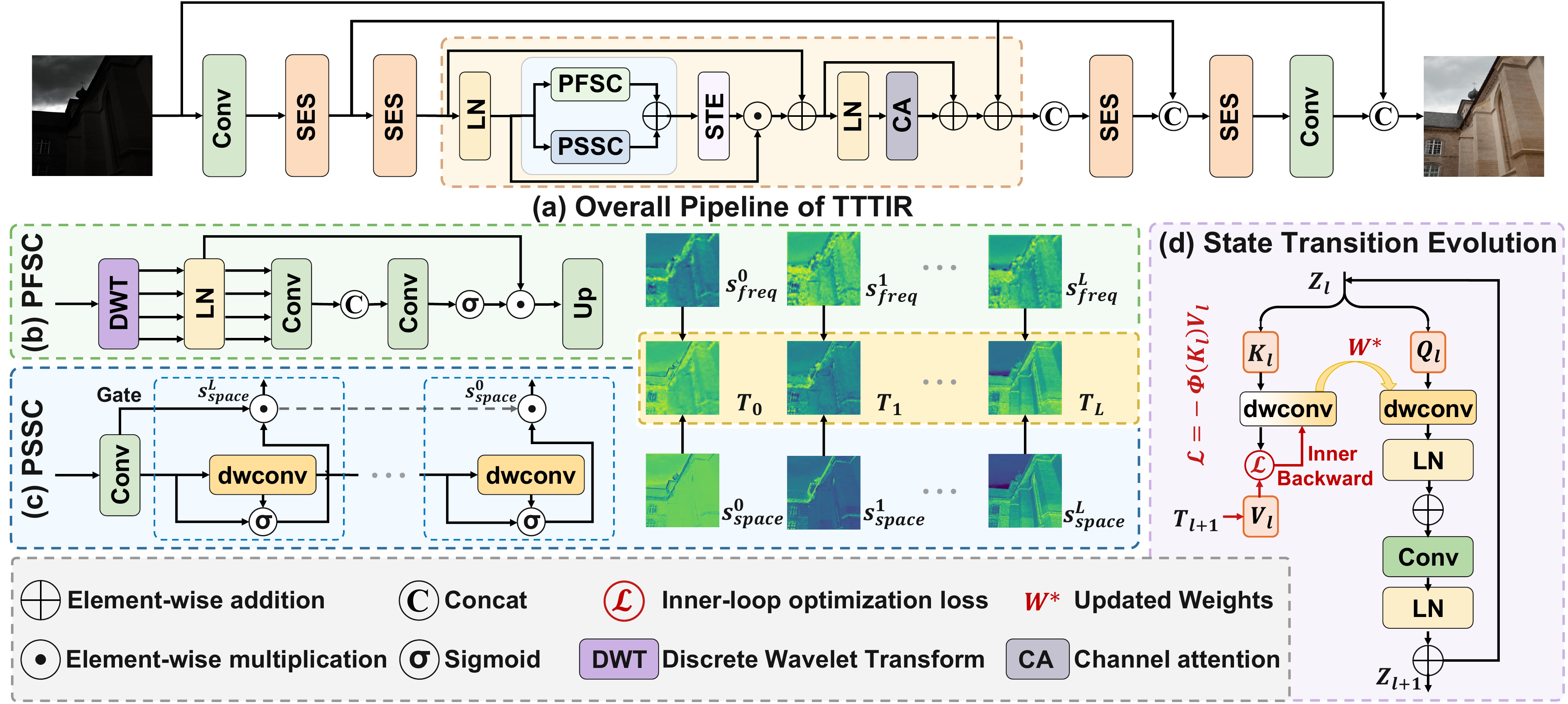}
    \caption{Framework of the proposed \modelname{}. 
    (a) Overall pipeline, where cascaded State Evolution Stages (SESs) progressively restore degraded features through progressive state evolution. 
    (b)-(c) The Progressive State Sequence Generation module constructs complementary frequency and spatial state sequences using PFSC and PSSC, respectively, which are aggregated into progressive target states. 
    (d) The State Transition Evolution (STE) module performs restoration-oriented test-time training by adaptively updating transition parameters to evolve the current state toward the next target state.}
    \label{fig:arc}
\end{figure*}
\subsection{Overall Framework}
\label{subsec:overview}

The architecture of \textbf{\modelname{}} is illustrated in \Cref{fig:arc}, where an encoder--decoder architecture is adopted as the backbone. Within each State Evolution Stage (SES), the input features are first processed by the Progressive State Sequence Generation (PSSG) module. By jointly exploiting spatial and frequency features, PSSG constructs a target state sequence \(\mathcal{T}\), providing progressive guidance for subsequent state evolution. In the State Transition Evolution (STE) module, based on the current restoration state \(Z_l\), the transition parameters are updated through a restoration-oriented inner-loop optimization process, with the next-level target state \(T_{l+1}\) serving as supervision. This enables the learning of adaptive state transition patterns during the restoration process. The transition parameters optimized through inner-loop adaptation are subsequently applied to the outer branch, and the evolved feature state \(Z_{l+1}\) is obtained through residual fusion with the current state. Finally, the Channel Attention enhances the features, producing the output \(Y^i\) of the SES.

\subsection{Progressive State Sequence Generation}
\label{subsec:sc}

Image restoration evolves from global structure recovery to local detail refinement. Accordingly, we propose PSSG to construct restoration-oriented state sequences by modeling spatial and frequency states, where \(\mathcal{S}_{space}\) captures structural information and \(\mathcal{S}_{freq}\) extracts complementary frequency features through adaptive spectral recombination. The two states are fused to generate the target state sequence \(\mathcal{T}\) for progressive restoration.

\noindent \textbf{Progressive Spatial State Constructor.} \quad
\label{subsubsec:PSSC} As shown in \mbox{\Cref{fig:arc}(c)}, PSSC constructs a sequence of spatial states by progressively enlarging the receptive field and extracting level-specific representations. Through recursive state evolution, it refines spatial contexts from global structural recovery to local texture reconstruction, producing complementary representations for restoration stages.

Given an input feature \(X \in \mathbb{R}^{B \times C \times H \times W}\), PSSC first applies a \(1 \times 1\) convolution to project it into an initial contextual state and a set of progressive gating maps:
\begin{equation}
[C_0,G_0,\dots,G_L]=\mathrm{Conv}_{1\times1}(X),
\end{equation}
where \(C_0\in\mathbb{R}^{B\times C\times H\times W}\) denotes the initial contextual representation, and \(G_l\in\mathbb{R}^{B\times1\times H\times W}\) represents the gating map at the \(l\)-th level. These gating maps are generated from the input features to preserve degradation-related spatial priors throughout the state evolution process.

At each level \(l\), the current contextual state \(C_l\) is first processed by a depth-wise convolution to obtain the enhanced representation \(\widetilde{C}_l\). The contextual state is then recursively updated through residual aggregation:
\begin{gather}
\widetilde{C}_{l} = \mathrm{DWConv}_{l}(C_l), \\
C_{l+1} = \gamma_l\odot C_l+\widetilde{C}_{l}.
\end{gather}
where \(\gamma_l \in \mathbb{R}^{1\times C\times1\times1}\) is a learnable scaling factor that balances the accumulated contextual information and newly aggregated features. Through this recursive evolution, the receptive field progressively expands from local regions to broader contextual regions. The level-wise spatial state is generated by gating the evolved representation:

\begin{equation}
s_l=C_{l+1}\odot G_l.
\end{equation}

The generated state sequence is reversed to align with the coarse-to-fine restoration trajectory:
\begin{equation}
\mathcal{S}_{space}=\{s_{space}^{0},s_{space}^{1},\dots,s_{space}^{L}\},
\quad s_{space}^{l}=s_{L-l}.
\end{equation}
After reordering, early states preserve global structure, while later states refine local textures.

\noindent \textbf{Progressive Frequency State Constructor} \quad
\label{subsubsec:PFSC} To capture complementary frequency details, we propose the PFSC (\mbox{\Cref{fig:arc}(b)}), which leverages frequency decomposition and adaptive band recombination to generate diverse progressive spectral states for subsequent evolution.

Given an input feature \(X\), PFSC applies DWT to obtain four frequency sub-bands, which are encoded by a shared depth-wise separable convolutional encoder:
\begin{equation}
\widetilde{B}^{m}=E(\mathrm{Norm}(\mathrm{DWT}(X)_m)).
\end{equation}
where \(m\) denotes the sub-band index with \(m\in\{LL,LH,HL,HH\}\). Here, \(LL\) captures structural information, while \(LH\), \(HL\), and \(HH\) encode high-frequency details. The frequency features are adaptively fused with level-specific weights to generate frequency states:
\begin{gather}
W_l^m = \mathrm{Softmax}(\mathrm{Conv}([\widetilde{B}^{m}])), \\
s_{freq}^{l}
=
\mathrm{Conv}
\left(
\sum_m W_l^m\odot \widetilde{B}^{m}
\right).
\end{gather}

By adaptively adjusting the contribution of different frequency bands, PFSC constructs a progressive frequency state sequence with diverse spectral representations:
\begin{equation}
\mathcal{S}_{freq}
=
\{s_{freq}^{0},s_{freq}^{1},\dots,s_{freq}^{L}\},
\end{equation}
where lower-level states emphasize low-frequency structural recovery, while higher-level states enhance high-frequency details, following the coarse-to-fine trajectory.

Finally, the spatial state sequence \(\mathcal{S}_{space}\) and frequency state sequence \(\mathcal{S}_{freq}\) are progressively aggregated to generate the target state sequence for subsequent evolution:
\begin{equation}
\mathcal{T}=\{T_0,T_1,\dots,T_L\},
\end{equation}
where $ T_l=\alpha_l s_{space}^{l}+(1-\alpha_l)s_{freq}^{l}.$
Here, \(\alpha_l\) controls the contribution of spatial and frequency states at each evolution level. The resulting target states provide progressive guidance for the subsequent STE module.

\subsection{State Transition Evolution}
\label{subsec:ste}

Conventional Test-Time Training (TTT) adapts a lightweight model through an inner-loop optimization process by constructing key-value pairs:
\begin{equation}
\hat{V} = f_{W}(K),\quad W \gets W - \eta \cdot \frac{\partial\mathcal{L}(\hat{V},V)}{\partial W},
\end{equation}
where \(K\) and \(V\) are derived from the input feature. However, such a formulation mainly focuses on contextual reconstruction and does not model the progressive evolution trajectory required by image restoration. Under complex degradations, restoration requires a continuous transition from coarse degradation suppression to fine-grained detail reconstruction.

In \Cref{fig:arc}(d), we reformulate TTT as a restoration-oriented
state transition process, where the state evolves toward a target state.
We initialize the restoration state as \(Z_0=T_0\).
Given the restoration state \(Z_l\) and the next target state
\(T_{l+1}\), we construct the query-key-value representations, adapt
the transition operator through inner-loop optimization, and apply the
operator to the query:
\begin{equation}
Q_l, K_l=\mathrm{MLP}_{in}(Z_l),
\quad
V_l=\operatorname{MLP}_{\mathrm{tar}}
\left(
\mathbf{T}_{l+1}
\right).
\end{equation}
\begin{equation}
\hat{V}_{l} = \Phi_{W_{l}}(K_l),
\quad
W_{l+1} = W_{l}-\eta\cdot
\frac{\partial\mathcal{L}(\hat{V}_{l},V_l)}
{\partial W_l},
\end{equation}
\begin{equation}
\hat{Q}_{l}=\Phi_{W_{l+1}}(Q_l).
\end{equation}
Here, \(Q_l\) and \(K_l\) are derived from the state
\(Z_l\) to capture transition patterns, while \(V_l\) is generated
from the target state \(T_{l+1}\) to provide guidance.
\(\Phi\) denotes a \(3\times3\) depth-wise convolutional
transition operator with learnable initialization \(W_0\).

Following~\cite{han2026vit}, we use the dot-product loss for inner-loop optimization, where \(N=HW\) is the number of positions and \(d_e=9\) the equivalent dimension of the \(3\times3\) DWConv:
\begin{equation}
\begin{aligned}
\mathcal{L}(\hat{V}_l,V_l)
&= \mathcal{L}(\Phi_{W_l}(K_l),V_l) \\
&= -\frac{1}{N\sqrt{d_e}}
\sum_{p=1}^{N}
\left\langle
\Phi_{W_l}(K_l)_p,V_{l,p}
\right\rangle,
\end{aligned}
\end{equation}

Different from conventional TTT that adapts parameters for context encoding, STE adapts the fast weights of \(\Phi\) to model restoration-specific transitions from the current state toward the desired progressive target state.

The evolved state is then refined through residual fusion:
\begin{gather}
R_l=
\mathrm{Norm}(\hat{Q})+Z_l, \\
Z_{l+1}
=
\mathrm{Conv}(R_l)+R_l .
\end{gather}

By updating the transition parameters according to the degradation condition, STE enables restoration state evolution: $Z_0\rightarrow Z_1\rightarrow\dots\rightarrow Z_L$ .
During this process, early states focus on degradation suppression and global structural recovery, while later states enhance texture reconstruction and fine-grained details.

Therefore, STE reformulates the conventional TTT objective from context reconstruction $K\rightarrow V$ into a restoration-oriented state transition $Z_l\rightarrow T_{l+1}$,
enabling adaptive learning of restoration trajectories from coarse degradation removal to fine-grained structural reconstruction.

\begin{table*}[t]
\centering
\caption{Quantitative comparison on LOL-v1, LOL-v2-Real, and LOL-v2-Synthetic datasets. The best and second-best results are highlighted in \textbf{bold} and \underline{underlined}, respectively. All results are obtained without using the GT-Mean strategy.}
\label{tab:lol_comparison}
\resizebox{\textwidth}{!}{%
\begin{tabular}{llcccccccc}
\toprule[0.15em]
\multirow{2}{*}{\textbf{Methods}} & \multirow{2}{*}{\textbf{Category}} 
 & \multicolumn{2}{c}{\textbf{LOL-v1}} 
 & \multicolumn{2}{c}{\textbf{LOL-v2-Real}} 
 & \multicolumn{2}{c}{\textbf{LOL-v2-Syn}} 
 & \multirow{2}{*}{\textbf{Param (M)}} & \multirow{2}{*}{\textbf{FLOPs (G)}} \\
 & & PSNR $\uparrow$ & SSIM $\uparrow$ & PSNR $\uparrow$ & SSIM $\uparrow$ & PSNR $\uparrow$ & SSIM $\uparrow$ & & \\
\midrule[0.15em]
Kind \cite{zhang2019kindling} & CNN-based & 20.87 & 0.7995 & 17.54 & 0.6695 & 22.62 & 0.9041 & 8.02 & 34.99 \\
Kind++ \cite{zhang2021beyond} & CNN-based & 17.97 & 0.8042 & 19.08 & 0.8176 & 21.17 & 0.8814 & 8.27 & 2970.5 \\
FourLLIE \cite{wang2023fourllie}& Frequency-based & 20.99 & 0.8071 & 23.45 & 0.8450 & 24.65 & 0.9192 & 0.12 & 4.07 \\
UHDFour \cite{li2023embedding} & Frequency-based & 22.89 & 0.8147 & 27.27 & 0.8579 & 23.64 & 0.8998 & 17.54 & 4.78 \\
DMFourLLIE \cite{zhang2024dmfourllie} & Frequency-based & 22.98 & 0.8273 & 26.40 & 0.8765 & \underline{25.74} & 0.9308 & 0.41 & 1.70 \\
FSFNet \cite{song2026low} & Frequency-based & 22.91 & \textbf{0.8600} & - & - & 25.30 & 0.8500 & 1.48 & 23.87 \\
Retinexformer \cite{cai2023retinexformer} & Transformer-based & 22.71 & 0.8177 & 24.55 & 0.8434 & 25.67 & 0.9295 & 1.61 & 15.57 \\
Wave-Mamba \cite{zou2024wave} & Mamba-based & 22.76 & 0.8419 & 27.87 & 0.8935 & 24.69 & 0.9271 & 1.26 & 7.22 \\
RetinexMamba \cite{bai2024retinexmamba} & Mamba-based & 23.15 & 0.8210 & 27.31 & 0.8667 & \textbf{25.89} & 0.9346 & 3.59 & 34.76 \\
CWNet \cite{zhang2025cwnet} & Mamba+Transformer & \underline{23.60} & 0.8496 & \underline{27.39} & \underline{0.9005} & 25.50 & \underline{0.9362} & 1.23 & 11.30 \\
MSHCDI-Net \cite{chen2026enhancing} & CNN+Transformer & 23.45 & 0.8480 & - & - & 23.74 & 0.9100 & 21.74 & 72.12 \\
\midrule
\modelname{} \textbf{(Ours)} & TTT-based & \textbf{23.76} & \underline{0.8556} & \textbf{30.78} & \textbf{0.9070} & 25.64 & \textbf{0.9400} & 0.787 & 18.71 \\
\bottomrule[0.15em]
\end{tabular}%
}
\end{table*}

\begin{figure*}[t!]
    \centering
    \includegraphics[width=.95\textwidth]{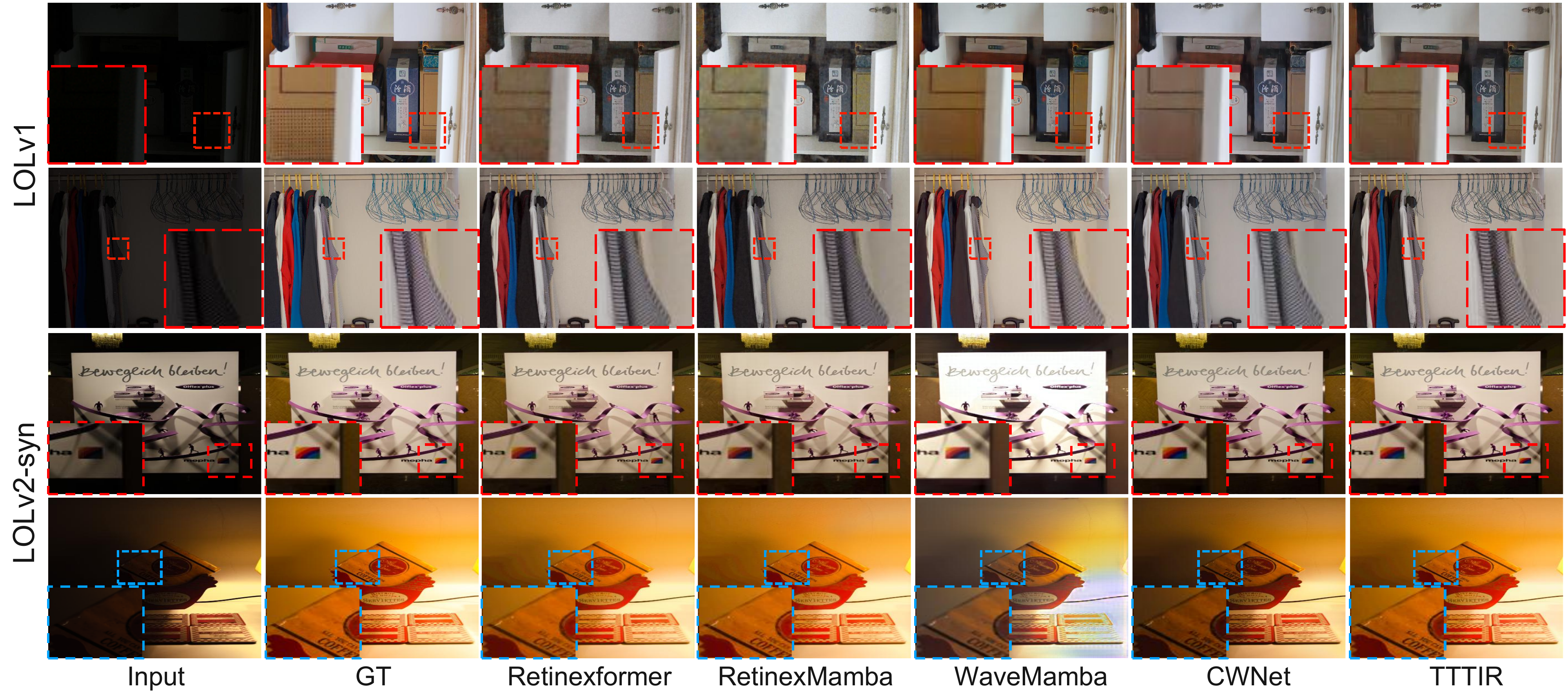}
    \caption{Visual comparison of different methods on the LOLv1 and LOLv2-syn datasets.}
    \label{fig:lol}
\end{figure*}

\section{Experiments}
\label{sec:experiments}
In this section, we evaluate \modelname{} on 12 benchmark datasets across three restoration tasks, including low-light enhancement, rain removal, and image dehazing.

\subsection{Datasets and Experimental Setting}
\label{subsec:setup}

\noindent \textbf{Datasets} \quad
\label{subsubsec:dataset_setup} For low-light image enhancement, we adopt the LOL-v1 , LOLv2-real and LOLv2-synthetic datasets, following previous work \cite{zhang2025cwnet}. For LOLv2-Real evaluation, we use the model trained on LOLv1 to demonstrate cross-dataset generalization. For rain streak removal, we train our model on Rain13K and evaluate it on Test100, Rain100H, Rain100L, Test1200, and Test2800 . For raindrop removal, we conduct both training and evaluation on the Raindrop-A and Raindrop-B datasets. For image dehazing, we utilize the RESIDE-6K and Haze4K datasets.

\noindent \textbf{Implementation Details and Evaluation}
\label{subsubsec:details_evaluation} \quad 
\modelname{} adopts a hierarchical encoder-decoder framework with a unified state evolution mechanism. 
Task-specific encoder-decoder configurations are employed for different restoration tasks, while the state evolution design remains unchanged. 
Detailed configurations are provided in the supplementary material. 
The embedding dimension is set to 32, with \(L=2\) and \(\alpha=0.5\). 
We employ the AdamW optimizer with \((\beta_1,\beta_2)=(0.9,0.999)\). 
Data augmentation includes random cropping and horizontal flipping. 
The learning rate follows a cosine annealing schedule with task-specific learning rates. 
The training objective combines an L1 reconstruction loss and a frequency-domain loss. 
All experiments are conducted on NVIDIA RTX 3080 Ti GPUs.
Following previous works, deraining results are reported in the YCbCr color space, while the remaining tasks are evaluated in the RGB color space.

\begin{table*}[t]
\centering
\caption{Quantitative comparison of \modelname{} with state-of-the-art methods on five synthetic rain streak removal benchmarks. The best and second-best results are highlighted in \textbf{bold} and \underline{underlined}, respectively.}
\label{tab:derain_streak}
\resizebox{\linewidth}{!}{
\begin{tabular}{l *{10}{c} || c c }
\toprule[0.15em]
 \multirow{2}{*}{\textbf{Methods}}& \multicolumn{2}{c}{\textbf{Test100}} & \multicolumn{2}{c}{\textbf{Rain100H}} & \multicolumn{2}{c}{\textbf{Rain100L}} & \multicolumn{2}{c}{\textbf{Test2800}} & \multicolumn{2}{c||}{\textbf{Test1200}} & \multicolumn{2}{c}{\textbf{Average}} \\
 & PSNR~$\textcolor{black}{\uparrow}$ & SSIM~$\textcolor{black}{\uparrow}$  & PSNR~$\textcolor{black}{\uparrow}$ & SSIM~$\textcolor{black}{\uparrow}$ & PSNR~$\textcolor{black}{\uparrow}$ & SSIM~$\textcolor{black}{\uparrow}$ & PSNR~$\textcolor{black}{\uparrow}$ & SSIM~$\textcolor{black}{\uparrow}$ & PSNR~$\textcolor{black}{\uparrow}$ & SSIM~$\textcolor{black}{\uparrow}$ & PSNR~$\textcolor{black}{\uparrow}$ & SSIM~$\textcolor{black}{\uparrow}$  \\
\midrule[0.15em]
DANet \cite{jiang2022danet}& 23.96 & 0.839 & 23.00 & 0.791 & 29.51 & 0.906 & 30.32 & 0.903 & 29.99 & 0.888 & 27.36 & 0.865 \\
Uformer \cite{wang2022uformer}& 23.87 & 0.815 & 22.43 & 0.700 & 28.39 & 0.883 & 29.71 & 0.886 & 28.65 & 0.856 & 26.61 & 0.828 \\
ALformer \cite{jiang2022magic}& 24.41 & 0.844 & 25.10 & 0.807 & 29.39 & 0.903 & 31.36 & 0.916 & 30.40 & 0.897 & 28.13 & 0.874 \\
NAFNet \cite{chen2022simple}& 25.75 & 0.845 & 26.76 & 0.813 & 31.27 & 0.925 & 31.71 & 0.918 & 30.62 & 0.892 & 29.22 & 0.879 \\
MFDNet \cite{wang2023multi}& 25.90 & 0.870 & 27.06 & 0.850 & 32.76 & 0.944 & 31.92 & 0.925 & 31.15 & 0.909 & 29.76 & 0.899 \\
HCT-FFN \cite{chen2023hybrid}& 24.86 & 0.847 & 26.70 & 0.819 & 29.94 & 0.906 & 31.46 & 0.915 & 31.23 & 0.901 & 28.84 & 0.878 \\
DRSformer \cite{chen2023learning}& 27.86 & 0.885 & 28.16 & 0.864 & 34.79 & 0.954 & 32.80 & 0.931 & 30.99 & 0.906 & 30.92 & 0.908 \\
ChaIR \cite{cui2023exploring}& 28.19 & 0.879 & 28.69 & 0.862 & 34.52 & 0.953 & 32.85 & 0.931 & 31.30 & 0.903 & 31.11 & 0.906 \\
FSNet \cite{cui2023image} & 27.95 & 0.884 & 28.70 & 0.860 & 34.10 & 0.952 & 32.68 & 0.931 & 31.26 & \underline{0.910} & 30.94 & 0.908\\
IRNeXT \cite{cui2023irnext}& 25.80 & 0.860 & 27.22 & 0.833 & 31.65 & 0.931 & 30.53 & 0.917 & 29.02 & 0.898 & 28.85 & 0.888 \\
OKNet \cite{cui2024omni}& 25.43 & 0.858 & 24.01 & 0.804 & 31.19 & 0.928 & 29.32 & 0.911 & 27.56 & 0.886 & 27.50 & 0.877 \\
AST \cite{zhou2024adapt}& 26.07 & 0.859 & 27.40 & 0.833 & 32.03 & 0.932 & 31.65 & 0.921 & 30.69 & 0.897 & 29.57 & 0.889 \\
SFHformer \cite{jiang2024fast}& 25.67 & 0.856 & 27.25 & 0.832 & 32.97 & 0.944 & 32.27 & 0.925 & 31.50 & 0.904 & 29.94 & 0.892 \\
Nerd-rain \citep{chen2024bidirectional}& 27.16 & 0.869 & 28.07 & 0.838 & 33.72 & 0.949 & 32.63 & 0.927 & 30.45 & 0.890 & 30.41 & 0.895 \\
AdaIR \cite{cui2025adair} & 28.64 & 0.889 & 29.48 & 0.871 & 35.84 & 0.962 & 32.70 & 0.930 & 30.58 & 0.907 & 31.45 & 0.912\\
CPRAformer \cite{zou2025cross} & \textbf{29.65} & \textbf{0.895} & \underline{29.68} & \underline{0.875} & \underline{35.98} & \underline{0.964} & \underline{33.00} & \textbf{0.933} & \textbf{31.52} & \textbf{0.913} & \underline{31.97} & \underline{0.916} \\
\midrule
\textbf{\modelname{}} (Ours) & \underline{29.57} & \underline{0.889} & \textbf{30.51} & \textbf{0.894} & \textbf{37.17} & \textbf{0.973} & \textbf{33.18} & \underline{0.931} & \underline{31.33} & 0.905 & \textbf{32.35} & \textbf{0.918}\\
\bottomrule[0.15em]
\end{tabular}
}
\end{table*}
\begin{figure*}[t!]
    \centering
    \includegraphics[width=.95\textwidth]{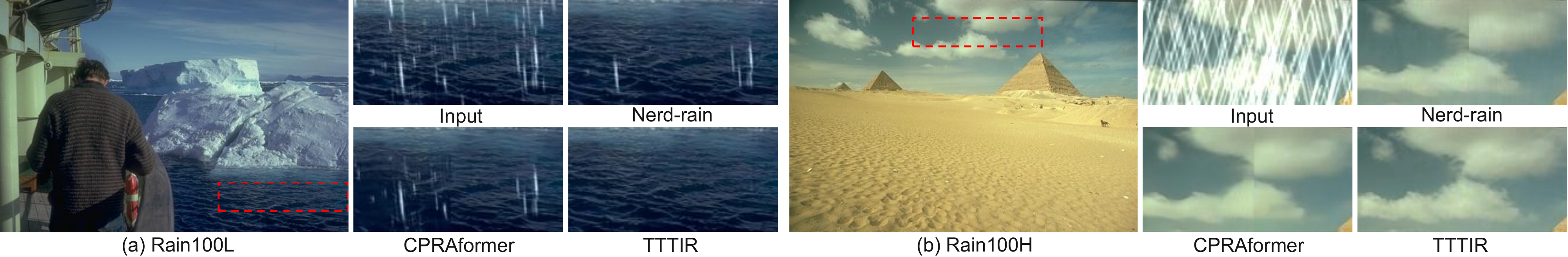}
    \caption{Visual comparison with Nerd-Rain and CPRAformer  on Rain100L and Rain100H.}
    \label{fig:rain100}
\end{figure*}

\begin{table}[t!]
\centering
\caption{Quantitative comparison of \modelname{} with state-of-the-art methods on the Raindrop dataset. The best and second-best results are highlighted in \textbf{bold} and \underline{underlined}, respectively.}
\label{tab:raindrop_removal}
\resizebox{\linewidth}{!}{
\begin{tabular}{l c c c c || c c}
\toprule[0.15em]
 \multirow{2}{*}{\textbf{Methods}}& \multicolumn{2}{c}{\textbf{Raindrop-A}} 
 & \multicolumn{2}{c||}{\textbf{Raindrop-B}}
 & \multicolumn{2}{c}{\textbf{Average}} \\

& \textbf{PSNR} $\uparrow$ & \textbf{SSIM} $\uparrow$
& \textbf{PSNR} $\uparrow$ & \textbf{SSIM} $\uparrow$
& \textbf{PSNR} $\uparrow$ & \textbf{SSIM} $\uparrow$ \\
\midrule[0.15em]
DANet \cite{jiang2022danet} & 29.54 & 0.914 & 25.28 & 0.812 & 27.41 & 0.863 \\
Uformer \cite{wang2022uformer} & 28.99 & 0.903 & 25.02 & 0.803 & 27.00 & 0.853 \\
ALformer \cite{jiang2022magic} & 29.11 & 0.911 & 25.11 & 0.809 & 27.11 & 0.860 \\
NAFNet \cite{chen2022simple} & 29.81 & 0.907 & 25.33 & 0.806 & 27.57 & 0.857 \\
MFDNet \cite{wang2023multi} & 28.57 & 0.882 & 24.53 & 0.766 & 26.55 & 0.824 \\
HCT-FFN \cite{chen2023hybrid} & 28.09 & 0.891 & 24.48 & 0.791 & 26.29 & 0.841 \\
DRSformer \cite{chen2023learning} & 30.83 & 0.923 & 25.86 & 0.819 & 28.34 & 0.871 \\
ChaIR \cite{cui2023exploring} & 30.88 & 0.925 & 25.84 & 0.820 & 28.36 & 0.873 \\
IRNeXT \cite{cui2023irnext} & 30.69 & 0.924 & 25.79 & 0.819 & 28.24 & 0.871 \\
OKNet \cite{cui2024omni} & 30.39 & 0.924 & 25.65 & 0.818 & 28.02 & 0.871 \\
SFHformer \cite{jiang2024fast} & 23.09 & 0.869 & 21.23 & 0.772 & 22.16 & 0.821 \\
Nerd-Rain \cite{chen2024bidirectional} & 30.96 & 0.924 & 25.96 & 0.819 & 28.46 & 0.872 \\
MSDT \cite{chen2024rethinking} & 30.85 & 0.922 & 25.89 & 0.818 & 28.37 & 0.870 \\
FSNet \cite{cui2023image} & 30.83 & 0.925 & 25.99 & 0.819 & 28.41 & 0.872 \\
AdaIR \cite{cui2025adair} & 30.99 & 0.924 & 25.97 & 0.817 & 28.48 & 0.871 \\
CPRAformer \cite{zou2025cross} & \underline{31.19} & \underline{0.926} & \underline{26.01} & \underline{0.821} & \underline{28.60} & \underline{0.874} \\
\midrule
\textbf{\modelname{}} (Ours) 
& \textbf{32.40} & \textbf{0.940} 
& \textbf{26.75} & \textbf{0.832}
& \textbf{29.57} & \textbf{0.886} \\

\bottomrule[0.15em]
\end{tabular}
}
\end{table}
\begin{table}[htbp]
\centering
\caption{Quantitative comparison of different methods on RESIDE-6K and Haze4K datasets. The best and second-best results are highlighted in \textbf{bold} and \underline{underlined}, respectively.}
\label{tab:dehaze}
\resizebox{\linewidth}{!}{
\begin{tabular}{l c c c c || c c}
\toprule[0.15em]
 \multirow{2}{*}{\textbf{Methods}}& \multicolumn{2}{c}{\textbf{RESIDE-6K}} 
 & \multicolumn{2}{c||}{\textbf{Haze4K}}
 & \multicolumn{2}{c}{\textbf{Average}} \\
& PSNR~$\uparrow$ & SSIM~$\uparrow$
& PSNR~$\uparrow$ & SSIM~$\uparrow$
& PSNR~$\uparrow$ & SSIM~$\uparrow$ \\
\midrule[0.15em]

Uformer \cite{wang2022uformer} 
& 26.29 & 0.925 
& 26.43 & 0.937 
& 26.36 & 0.931 \\

LKD \cite{luo2023lkd}
& 25.42 & 0.925 
& 27.39 & 0.938 
& 26.41 & 0.932 \\

Dehazeformer \cite{song2023vision}
& 26.25 & 0.931 
& 27.45 & 0.946 
& 26.85 & 0.939 \\

MB-TaylorFormer \cite{qiu2023mb}
& 26.28 & 0.923 
& 26.34 & 0.933 
& 26.31 & 0.928 \\

MixDehazeNet \cite{lu2023mixdehazenet}
& 26.62 & 0.939 
& 27.34 & 0.945 
& 26.98 & 0.942 \\

DEANet \cite{chen2024dea}
& 26.61 & 0.932 
& 26.94 & 0.942 
& 26.78 & 0.937 \\

SFHformer \cite{jiang2024fast}
& 27.08 & 0.940 
& 26.92 & 0.941 
& 27.00 & 0.941 \\

CPRAformer \cite{zou2025cross}
& \textbf{27.70} & \underline{0.944}
& \underline{27.97} & \underline{0.952}
& \textbf{27.84} & \underline{0.948} \\
\midrule
\textbf{\modelname{}} (Ours) 
& \underline{27.54} & \textbf{0.951}
& \textbf{28.06} & \textbf{0.970}
& \underline{27.80} & \textbf{0.961} \\

\bottomrule[0.15em]
\end{tabular}
}
\end{table}
\subsection{Comparasion with State-of-The-Arts}
\label{subsec:sota}

\noindent \textbf{Low-Light Enhancement} \quad
\label{subsubsec:lle_sota}
The comparisons on LOL datasets are presented in \Cref{tab:lol_comparison}. 
\modelname{} achieves state-of-the-art performance on LOL-v2-Real with 30.78 dB PSNR and 0.9070 SSIM. It obtains the highest SSIM of 0.9400 on LOL-v2-Synthetic and ranks first in PSNR on LOL-v1. The visual comparisons in \Cref{fig:lol} validate its effectiveness in illumination and detail restoration.

Moreover, \modelname{} requires only 0.787M parameters and 18.71G FLOPs, making it more lightweight than most Transformer- and Mamba-based methods while maintaining competitive restoration performance.

\noindent \textbf{Rain Streak Removal} \quad
The results are reported in \Cref{tab:derain_streak}. 
\modelname{} achieves the highest average PSNR and SSIM of 32.35 dB and 0.918, surpassing CPRAformer by 0.38 dB and 0.002, respectively. 
On Rain100H and Rain100L, \modelname{} obtains 30.51 dB/0.894 SSIM and 37.17 dB/0.973 SSIM, respectively, demonstrating superior rain streak removal and structural preservation. 
On Test100 and Test2800, \modelname{} also achieves competitive performance. The qualitative comparisons are presented in \Cref{fig:rain100}.

\noindent \textbf{Raindrop Removal} \quad
The quantitative comparisons on the Raindrop dataset are reported in \Cref{tab:raindrop_removal}. 
\modelname{} achieves the best average performance with 29.57 dB PSNR and 0.886 SSIM, outperforming CPRAformer by 0.97 dB and 0.012, respectively. 
On Raindrop-A and Raindrop-B, \modelname{} consistently achieves the highest PSNR and SSIM, demonstrating its robustness in handling diverse raindrop patterns.

\begin{figure}[t]
    \centering
    \includegraphics[width=\linewidth]{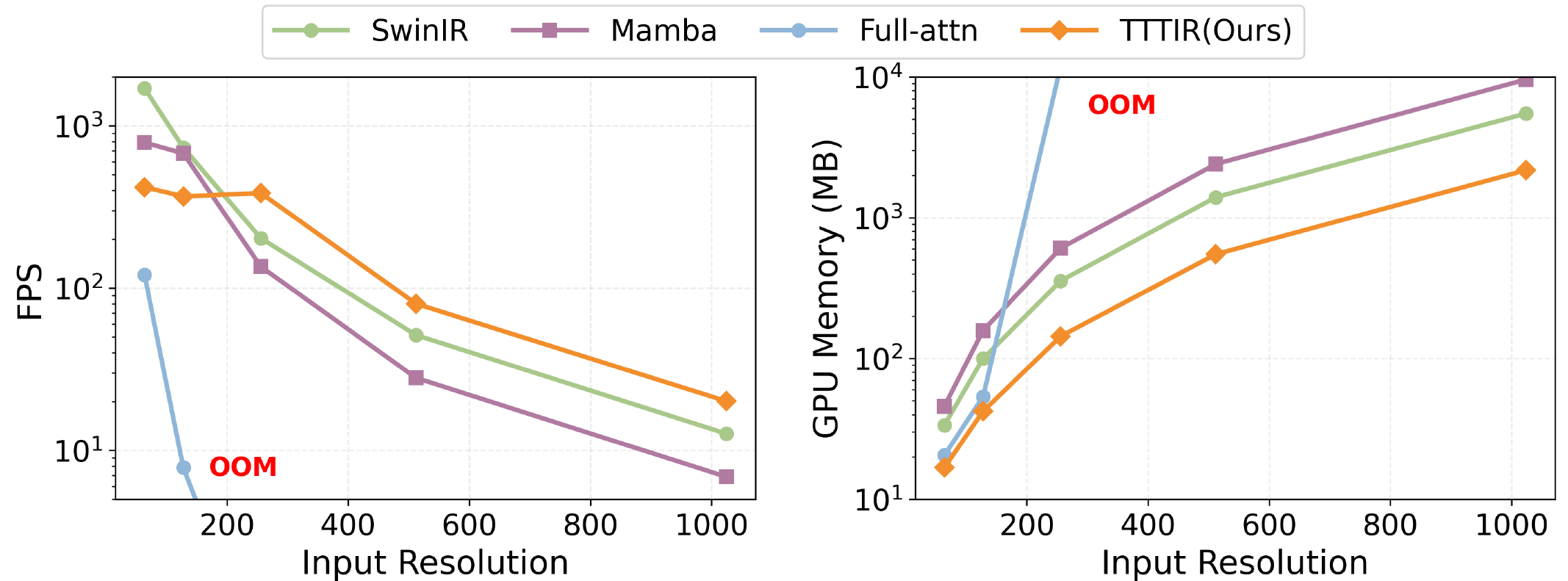}
    \caption{Computational complexity comparison with different input scales. \modelname{} achieves favorable scalability in both inference speed and GPU memory consumption.}
    \label{fig:efficiency}
\end{figure}

\noindent \textbf{Dehaze} \quad
\label{subsubsec:dehaze_sota}
We further evaluate \modelname{} on the image dehazing task. The quantitative comparisons are reported in \Cref{tab:dehaze}. On the RESIDE-6K dataset, \modelname{} achieves the highest SSIM score of 0.951. On the Haze4K dataset, \modelname{} achieves the best performance in terms of both PSNR and SSIM, reaching 28.06 dB and 0.970, respectively, outperforming all compared methods. The consistent improvements demonstrate that \modelname{} can effectively model complex haze degradation and reconstruct clear image content. This benefits from the proposed degradation-aware feature representation and progressive feature interaction mechanism, which enable the network to capture both global atmospheric information and local texture details during the restoration process.

\begin{figure}[t!]
    \centering
    \includegraphics[width=\linewidth]{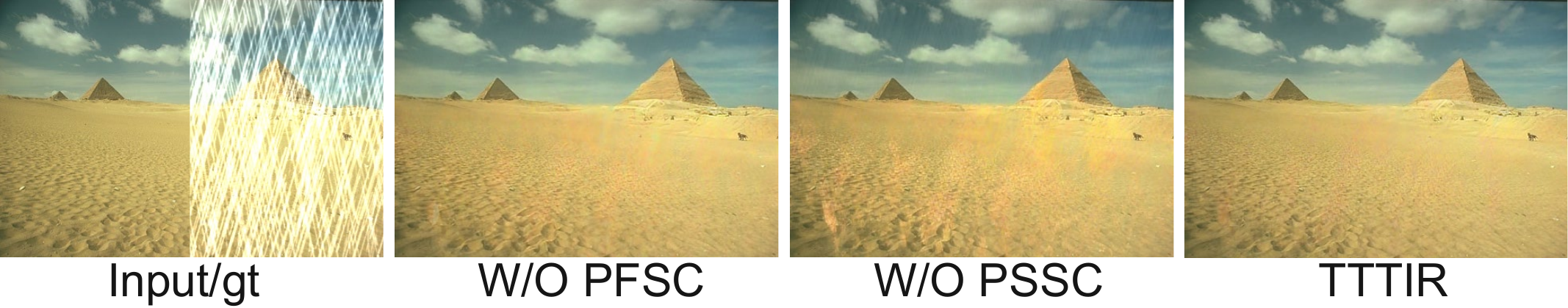}
    \caption{Qualitative ablation comparison of \modelname{} with w/o PSSC and w/o PFSC for rain removal.}
    \label{fig:abs}
\end{figure}

\noindent \textbf{Efficiency Analysis} \quad
\label{subsec:efficiency}
In addition to restoration quality, we evaluate the computational scalability of \modelname{} under different input resolutions. 
As shown in \Cref{fig:efficiency}, \modelname{} is compared with Transformer-based and Mamba-based architectures in terms of inference speed and GPU memory consumption. All compared models are configured with comparable parameter counts to ensure a fair efficiency comparison. With increasing resolution, full-attention models suffer from severe computational and memory overhead due to quadratic global interactions, resulting in rapid cost growth. SwinIR-based and Mamba-based methods exhibit similar scaling trends to \modelname{} but incur higher computational and memory costs under the same resolutions.
In contrast, \modelname{} keeps lower inference latency and memory overhead across resolutions. This advantage is attributed to the progressive state evolution mechanism, which avoids costly global interactions and progressively models restoration states through lightweight spatial-frequency transitions.

\subsection{Ablation Study}

\begin{table}[t]
\centering
\caption{Ablation study of different components in \modelname{}.}
\label{tab:ablation}
\resizebox{\linewidth}{!}{
\begin{tabular}{l c c c c || c c}
\toprule[0.15em]
\multirow{2}{*}{\textbf{Methods}}& \multicolumn{2}{c}{\textbf{Rain100H}}
& \multicolumn{2}{c||}{\textbf{Rain100L}}
& \multicolumn{2}{c}{\textbf{Complexity}} \\
& PSNR~$\uparrow$ & SSIM~$\uparrow$
& PSNR~$\uparrow$ & SSIM~$\uparrow$
& Params (M) & FLOPs (G) \\
\midrule[0.15em]

Baseline 
&28.06&0.8457
&34.66&0.9578
&1.15&12.37
\\

baseline+PSSG
&29.44&0.875
&36.34&0.968
&1.55&23.57
\\

baseline+STE
&29.35&0.878
&37.15&0.973
& 1.35 & 14.21
\\

w/o PSSC
&29.38&0.8761
&36.19&0.9669
&2.17&28.61
\\

w/o PFSC
&30.11&0.8866
&37.11&0.9719
&2.00&22.42
\\

w/ Current-State Target
&30.12&0.8863
&36.84&0.9715
&2.65&34.89
\\

\modelname{}
&\textbf{30.51}&\textbf{0.8940}
&\textbf{37.17}&\textbf{0.9730}
&2.29&31.11
\\

\bottomrule[0.15em]
\end{tabular}
 }
\end{table}
\label{sec:ablation}
\begin{table}[t]
\centering
\caption{Effect of progressive state number $L$ on restoration performance and computational cost. Default: L=2.}
\label{tab:ablation_L}
\resizebox{\linewidth}{!}{
\begin{tabular}{c c c c c || c c}
\toprule[0.15em]
& \multicolumn{2}{c}{\textbf{Rain100H}}
& \multicolumn{2}{c||}{\textbf{Rain100L}}
& \multicolumn{2}{c}{\textbf{Complexity}} \\
$L$
& PSNR~$\uparrow$ & SSIM~$\uparrow$
& PSNR~$\uparrow$ & SSIM~$\uparrow$
& Params (M) & FLOPs (G) \\
\midrule[0.15em]

1
& 29.91 & 0.883
& 36.68 & 0.971
& 1.92 & 24.36
\\

2
& 30.51 & \textbf{0.894}
& 37.17 & \textbf{0.973}
& 2.29 & 31.11
\\

3
& \textbf{30.55} & 0.893
& \textbf{37.43} & 0.972
& 2.66 & 37.85
\\

\bottomrule[0.15em]
\end{tabular}
}
\end{table}

We conduct ablations on Rain100L and Rain100H to evaluate \modelname{} components. 
PSNR, SSIM, Params, and FLOPs are used for evaluation. 
The baseline adopts a 3-stage encoder-decoder with CAB blocks and residual connections, with variants trained under identical settings.

\noindent\textbf{Effect of Progressive State Construction and Evolution}
\quad
We study three variants to evaluate the contributions of progressive
state construction, inner-loop evolution, and next-state guidance:
(1) \textbf{Baseline+PSSG}, which introduces progressive
spatial-frequency states without inner-loop transitions;
(2) \textbf{Baseline+STE}, which applies self-supervised inner-loop
evolution without PSSG, where both keys and values are derived from
the current feature; and
(3) \textbf{w/ Current-State Target}, which retains the
complete architecture but replaces the next-state target $T_{l+1}$
with the current-level target $T_l$:
\begin{equation}
\mathcal{L}_{\mathrm{cur}}
=
\mathcal{L}
\left(
\Phi_{\theta}(K(Z_l)),V(T_l)
\right).
\end{equation}
As shown in \Cref{tab:ablation}, Baseline+PSSG and Baseline+STE both
outperform the baseline, validating the effectiveness of progressive
state construction and inner-loop evolution, respectively.
Moreover, \modelname{} outperforms the Current-State Target variant,
showing that next-state prediction provides more effective guidance
for restoration.

\noindent \textbf{Effect of Spatial and Frequency State Constructors} \quad
To investigate the contribution of the proposed PSSC and PFSC, 
we design two ablative variants: 
(i) \textbf{w/o PSSC}, which removes PSSC to evaluate the importance of spatial structural priors; and 
(ii) \textbf{w/o PFSC}, which removes PFSC to assess the contribution of frequency-aware degradation priors.
As shown in \Cref{tab:ablation}, removing either PSSC or PFSC degrades performance, confirming the importance of both spatial and frequency information. The larger drop observed without PSSC highlights the key role of spatial structure, while PFSC provides complementary frequency cues. Qualitative results in \Cref{fig:abs} further validate the effectiveness of both modules.

\noindent \textbf{Effect of Progressive State Number \(L\)} \quad 
We investigate the impact of progressive state number $L$.  As shown in \Cref{tab:ablation_L}, increasing $L$ enables a refined restoration trajectory.  Increasing $L$ from $1$ to $2$ introduces a coarse-to-fine evolution paradigm, bringing significant PSNR gains by allowing the network to prioritize global structural recovery before refining local details. However, further increasing $L$ to $3$ yields only marginal improvements while incurring disproportionate computational overhead during the inner-loop optimization, indicating a diminishing return in state refinement. Therefore, to achieve an optimal trade-off between restoration efficacy and computational efficiency, we adopt $L=2$ as the default configuration.
\section{Conclusions}
\label{sec:conclusion}

In this work, we present \modelname{}, a Test-Time Training (TTT) framework that reformulates image restoration as an input-dependent, progressive state evolution process. Diverging from existing methods restricted by globally shared parameters, \modelname{} dynamically adapts to the degradation characteristics of each image instance. Specifically, we construct complementary spatial-frequency target states to define \emph{what} to recover, while introducing a restoration-oriented TTT objective that optimizes a lightweight transition operator via inner-loop updates to determine \emph{how} the feature evolution should be performed. By enabling explicit, instance-specific state transitions rather than static mappings, our framework effectively handles diverse real-world degradations. Extensive experiments across multiple benchmarks demonstrate that \modelname{} achieves superior restoration performance and favorable scalability over state-of-the-art approaches.
\bibliography{main}

\end{document}